\documentclass[letterpaper]{article}
\usepackage{spconf,amsmath,amsfonts,amssymb}
\usepackage{graphicx}
\usepackage[hidelinks]{hyperref}
\usepackage{subfig}
\usepackage{xcolor}
\usepackage{braket}
\usepackage{float}
\usepackage{booktabs}
\usepackage{bm}

\title{MIMO Array Calibration in Non-stationary Channels with Residual Surfaces and Slepian Spherical Harmonics}
\name{Oliver Kirkpatrick, Santiago Ozafrain, Christopher Gilliam, Beth Jelfs}
\address{School of Engineering, University of Birmingham, UK\\ opk323@student.bham.ac.uk, \{s.ozafrain, c.gilliam.1, b.jelfs\}@bham.ac.uk}

\begin{document}
\ninept 
\maketitle 
\begin{abstract}
The fundamental mechanism driving MIMO beamforming is the relative phases of signals departing the transmit array and arriving at the receive array. If a propagation channel affects all transmitted signals equally, the relative phases are a function of the directions of departure and arrival, as well as the transmit and receive hardware. In a non-stationary channel, the amplitudes and phases of arriving signals may vary significantly over time, making it infeasible to directly measure the influence of hardware. In this paper, we present a calibration method for achieving indirect measurement and compensation of hardware influences in non-stationary channels. Our method characterizes the patterns of array elements relative to a reference element and estimates these relative patterns, termed \textit{residual surfaces}, using a Slepian spherical harmonic basis. Using simulations, we demonstrate that our calibration method achieves beamforming gains that closely match theoretical optimums. Our results also show a reduction in the error in estimating the target direction, lower side lobes, and improve null-steering capabilities.
\end{abstract}


\section{Introduction}
\label{sec:Intro}
The design and use of multiple-input-multiple-output (MIMO) antenna arrays continues to be an ongoing area of research~\cite{molaei2025efficient,deng2022limited,xue2024optimized}. The benefits of a well-designed MIMO system include improved transmit beam pattern design, spatial diversity~\cite{molaei2025efficient}, interference rejection and anti-jamming~\cite{deng2022limited,xue2024optimized}, and clutter rejection~\cite{abramovich2015aperiodicwaveforms}. The extent of these advantages depends greatly on the geometry of the transmit and receive arrays~\cite{7060413}, both in their design and in their manufacture and deployment. An operational antenna system will require gain pattern measurements to compensate for hardware-induced perturbations to signal amplitude and phase on transmit and receive~\cite{1455209}. In MIMO radar, this process is still an open question~\cite{bergin2018mimo}, especially when operating in environments with complex non-stationary channels, which can significantly change the properties of the received signal. This can be a particular problem in situations such as marine radar, where the characteristics of the clutter can vary significantly over time~\cite{katzin1957mechanisms}, or over-the-horizon radar, where the ionosphere causes significant variations in the signal strength and phase~\cite{fabrizio2013high}.

Traditional methods for sampling antenna gain patterns make use of relocatable probe stations, turn tables, or controlled mobile targets. Measurements often take place inside an anechoic chamber, offering a tightly controlled environment and tailored propagation paths to accommodate many different frequencies and sampling methods~\cite{1455209,balanis2015antenna}. However, the spatial limitations of indoor test facilities physically limit the size of antennas that can be tested or are incompatible with the size of the far field; this is particularly problematic for high frequency (HF) antenna arrays that operate with wavelengths between 10 and 100 meters. In addition, the operating environment of the array can alter gain patterns, making anechoic chamber measurements redundant, particularly for non-stationary environments.

In contrast, outdoor measurements can be made in situ using mobile probe stations (ground-based or airborne~\cite{fabrizio2013high,balanis2015antenna}), photogrammetry~\cite{1455209}, or laser ranging techniques to check antenna geometry~\cite{payne1976new}. However, in situ measurements for physically large arrays can be time consuming and costly, and often require access to very large areas of land or accessible airspace for the mobile probe station~\cite{fabrizio2013high,balanis2015antenna}. Alternatively, known astronomical sources have been proposed~\cite{wu2014computationally,ollier2018robust}, addressing cost and probe station dependence; however, they are primarily applicable to receive arrays, ruling out bistatic systems (and hence MIMO systems). Orbital targets~\cite{fabrizio2013high} are an attractive option, but their use assumes that the target can be illuminated at the operational frequencies of the radar or that the calibration results can be mapped to the operational frequencies. Accordingly, our focus is on developing a calibration method that can be used in situ, with potentially large arrays or bistatic configurations, and without reliance on specific targets. 

In this paper, we propose a method for calibrating MIMO systems that are not well suited to traditional calibration methods. This method is based on measuring the relative phases between elements in an array and a designated reference element. These relative differences should remain the same for any given propagation path, hence making this method suitable for antenna array calibration in non-stationary channels, provided the propagation path is understood. We propose modeling the relative phases for an element using a continuous Slepian basis defined on the unit sphere and estimating the weights of the basis using radar measurements of known targets. We denote this model of the relative phase differences as the residual surface and estimate a surface per antenna element. These residual surfaces can then be integrated into the MIMO beamforming process to recover losses in beamforming performance associated with uncalibrated arrays. In addition, this method is also applicable to use with scatterers of opportunity with time-varying echo strength, or without a known radar cross section.

This paper is structured as follows: \autoref{sec:mimo-model} outlines MIMO beamforming and imperfect antenna gain patterns before deriving the residual surface for the perturbed beamforming model; \autoref{sec:residual-surfaces} lays out a method for constructing band-limited Slepian basis functions concentrated to the useful field of view of the radar, and how to estimate weights for these basis function from measurements; \autoref{sec:performance-analysis} demonstrates through Monte-Carlo simulations that the weights of the basis functions can be reliably estimated in the presence of noise, and that the corresponding functions improve beam pattern qualities; finally, \autoref{sec:conclusion} concludes the paper.

\section{Problem Formulation}
\label{sec:mimo-model}
\begin{figure}
    \centering
    \includegraphics[width=0.6\linewidth]{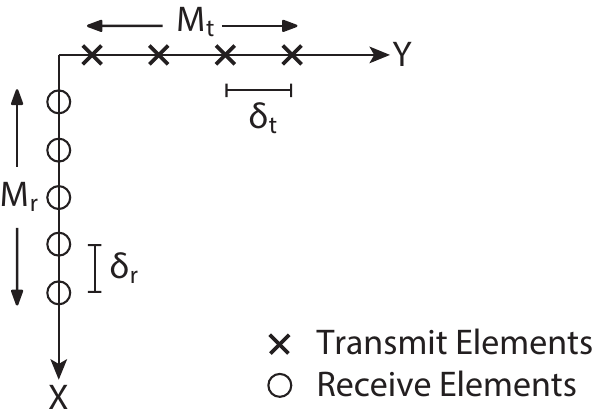}
    \caption{Geometry of orthogonal and co-located uniform linear transmit and receive arrays.}
    \label{fig:array-geometry}
\end{figure}
Consider a MIMO system consisting of two co-located and orthogonal linear arrays: a transmit array of $M_{t}$ elements and a receive array of $M_{r}$ elements, as depicted in \autoref{fig:array-geometry}. Elements within each linear array are spaced a half wavelength apart, $\delta_{t}=\delta_{r}=\lambda/2$, where $\lambda$ is the wavelength corresponding to the operating frequency of the array. We assume that at the receive array each element has a quadrature receiver recording IQ radar samples, thus providing both amplitude and phase information. In MIMO beamforming, direction finding is informed by relative phases between the transmit and receive elements. These relative phases can be defined in terms of the azimuth, $\vartheta$, and elevation, $\varphi$, of the departing or arriving signals~\cite{bergin2018mimo,qu2008performance}. This configuration gives rise to the transmit and receive steering vectors, $\mathbf{a}_{t}$ and $\mathbf{a}_{r}$, respectively, in the following form~\cite{bergin2018mimo}:
\begin{align}
    \mathbf{a}_{t}\left(\vartheta,\varphi\right)&=
    \begin{bmatrix}
        1\\
        \exp\left(-\pi jm\sin\vartheta\cos\varphi\right)\\
        \vdots\\
        \exp\left(-\pi j\left(M_{t}-1\right)\sin\vartheta\cos\varphi\right)
    \end{bmatrix},\label{eq:normal-beamforming}\\
    \mathbf{a}_{r}\left(\vartheta,\varphi\right)&=
    \begin{bmatrix}
        1\\
        \exp\left(-\pi jm\cos\vartheta\cos\varphi\right)\\
        \vdots\\
        \exp\left(-\pi j\left(M_{r}-1\right)\cos\vartheta\cos\varphi\right)
    \end{bmatrix},\label{eq:r_beamform}
\end{align}
where $m$ is the array index of the element. However, in practice, imperfections unique to each antenna in an array will induce phase and amplitude differences between channels, which are not characterized in~\eqref{eq:normal-beamforming} and~\eqref{eq:r_beamform}. These phases and amplitudes are a function of azimuth and elevation, and are referred to as the gain patterns $G_{t,m}\left(\vartheta,\varphi\right)$ and $G_{r,m}\left(\vartheta,\varphi\right)$ of the elements. We term the steering vectors inclusive of these complex-valued gain patterns as perturbed steering vectors, $\tilde{\mathbf{a}}_{t}$ and $\tilde{\mathbf{a}}_{r}$, defined as,
\begin{equation}\label{eq:new-steering-with-gain}
    \tilde{\mathbf{a}}_{t}\left(\vartheta,\varphi\right)=
    \begin{bmatrix}
        G_{t,0}\left(\vartheta,\varphi\right)\\
        G_{t,1}\left(\vartheta,\varphi\right)\exp\left(-\pi jm\sin\vartheta\cos\varphi\right)\\
        \vdots\\
        G_{t,M_{t}-1}\left(\vartheta,\varphi\right)\exp\left(-\pi j\left(M_{t}-1\right)\sin\vartheta\cos\varphi\right)
    \end{bmatrix},
\end{equation}
and
\begin{equation}\label{eq:new-steering-with-gain-r}
    \tilde{\mathbf{a}}_{r}\left(\vartheta,\varphi\right)=\\
    \begin{bmatrix}
        G_{r,0}\left(\vartheta,\varphi\right)\\
        G_{r,1}\left(\vartheta,\varphi\right)\exp\left(-\pi jm\cos\vartheta\cos\varphi\right)\\
        \vdots\\
        G_{r,M_{r}-1}\left(\vartheta,\varphi\right)\exp\left(-\pi j\left(M_{r}-1\right)\cos\vartheta\cos\varphi\right)
    \end{bmatrix}.
\end{equation}
We next define the combination of transmit and receive steering vectors from all elements as,
\begin{equation}\label{eq:kronecker-beamformer}
    \tilde{\mathbf{y}}=\tilde{\mathbf{a}}_{t}\!\left(\vartheta,\varphi\right)\otimes\tilde{\mathbf{a}}_{r}\!\left(\vartheta,\varphi\right)+\boldsymbol{\eta},
\end{equation}
where $\boldsymbol{\eta}$ is a vector of complex-valued additive white Gaussian noise and $\otimes$ denotes the Kronecker product. We assume uniform power across channels in $\tilde{\mathbf{a}}_{t}\otimes\tilde{\mathbf{a}}_{r}$, which is equivalent to assuming that the gain patterns are unit-magnitude complex values. 

At each receive element, the recorded IQ radar samples are processed to separate the transmit channels, and range and Doppler compression is performed on each transmit-receive pair. Peaks in the processed radar data can then be associated with known targets, or scatterers of opportunity. Therefore, providing the array geometry and departure/arrival directions corresponding to a target are known, we can remove the ideal steering vector from the measurement $\tilde{\mathbf{y}}$. Given a known $\vartheta$ and $\varphi$ we construct the diagonal matrix $\mathbf{A}=\operatorname{diag}\big(\mathbf{a}_{t}\left(\vartheta,\varphi\right)\otimes\mathbf{a}_{r}\left(\vartheta,\varphi\right)\big)$ and extract the complex gains (and effects due to the channel) from $\tilde{\mathbf{y}}$ as follows:
\begin{equation}\label{eq:gains-only}
    \mathbf{g}\left(\vartheta,\varphi\right)=\mathbf{A}^{\dagger}\tilde{\mathbf{y}} =\tilde{\mathbf{g}}_{t}\otimes\tilde{\mathbf{g}}_{r} + \bm{\varepsilon},
\end{equation}
where $(\cdot)^\dagger$ denotes the Hermitian conjugate, $\tilde{\mathbf{g}}_{k}=[G_{k,0}, G_{k,1},\allowbreak \ldots, G_{k,M_{k}-1}]$ with $k\in[t,r]$ representing either the transmit or receive array and $\bm{\varepsilon}=\mathbf{A}^{\dagger}\boldsymbol{\eta}$. The vector $\mathbf{g}\left(\vartheta,\varphi\right)$ is unit norm and represents different combinations of transmit and receive (complex) gains. 

Typically, $\tilde{\mathbf{g}}_{k}$ would be measured with traditional calibration methods, however, in a non-stationary propagation channel these methods are hampered by difficulty decoupling $G_{k,m}\left(\vartheta,\varphi\right)$ from time varying influences. Therefore, we introduce the \emph{residual surface} to address this via normalization relative to a designated reference channel, in this work we designate $m=0$ as the reference channel, giving:
\begin{equation}\label{eq:residual-vec}
    \mathbf{g}_{k}'\left(\vartheta,\varphi\right)=
    \begin{bmatrix}
        1\\
        G_{k,1}\left(\vartheta,\varphi\right)G_{k,0}\left(\vartheta,\varphi\right)^{*}\\
        \vdots\\
        G_{k,M_{k}-1}\left(\vartheta,\varphi\right)G_{k,0}\left(\vartheta,\varphi\right)^{*}
    \end{bmatrix},
\end{equation}
where $(\cdot)^*$ denotes the complex conjugate. The calibration process is now one of estimating~\eqref{eq:residual-vec} from processed radar data so that it can be sampled and applied during the beamforming process. In the following section, we will outline a method to estimate the residual surfaces for restricted fields of view of a MIMO array using targets or scatterers of opportunity with known directions of departure and arrival.

\section{Proposed Residual Surface Calibration Method}
\label{sec:residual-surfaces}

The aim of our proposed method is to estimate the functions $\Delta G_{k,m}\left(\varphi,\vartheta\right)=G_{k,m}\left(\varphi,\vartheta\right)G_{k,0}\left(\varphi,\vartheta\right)^{*}$ in~\eqref{eq:residual-vec} so that they can be used to perform compensation during the beamforming process. To achieve this aim, we propose representing the functions $\Delta G_{k,m}$ on an orthogonal basis defined on the unit sphere and then estimating the weights of this basis representation. The weights will be estimated using radar measurements corresponding to targets, or scatterers of opportunity, in known directions. However, in theory, the functions $\Delta G_{k,m}$ are defined on the complete unit sphere, whereas, in practice an operational radar is unlikely to be able to observe the entire sphere. Accordingly, we need to define the orthogonal basis on a subregion of the unit sphere and estimate the functions $\Delta G_{k,m}$ using observations of the surface obtained from the radar measurements. In the following we detail this procedure.

\subsection{Orthogonal basis functions tailored to a subregion}
For functions on the unit sphere, $\mathbb{S}^{2}$, like $\Delta G_{k,m}\left(\vartheta,\varphi\right)$, band-limited weighted spherical harmonic functions, are frequently used to approximate the continuous function~\cite{simons2015scalar}:
\begin{equation}
    f\left(\vartheta,\varphi\right)=\sum_{p=0}^{P}\sum_{q=-p}^{p}w_{pq}Y_{pq}\left(\vartheta,\varphi\right),
\end{equation}
where $Y_{pq}$ is the spherical harmonic of degree $p$ and order $q$, and $w_{pq}$ is the weight for that particular harmonic. For a maximum degree $P$, a total of $\left(P+1\right)^{2}$ harmonics are used. If we assume that the field of view of the radar is restricted to a subregion $\mathcal{R}$, then in this subregion, the spherical harmonics are not guaranteed to be orthogonal~\cite{simons2015scalar}. However, they can be used to construct a tailored basis for $\mathcal{R}$ using Slepian functions~\cite{simons2015scalar,bates2015use,wieczorek2007minimum}. 

A comprehensive derivation of Slepian spherical harmonic basis functions is beyond the scope of this paper, but can be found in~\cite{simons2015scalar}. In brief, to generate these tailored basis functions, we first calculate $\mathbf{Y}\in\mathbb{R}^{\left(P+1\right)^{2}\times N_b}$, where the elements of $\mathbf{Y}$ are the $Y_{pq}\left(\vartheta,\varphi\right)$ spherical harmonics evaluated at $N_b$ quadrature points inside $\mathcal{R}$, and the diagonal matrix $\mathbf{W}$, which holds the corresponding quadrature weights for each of the $N_b$ locations. If we assume $\mathbf{D}=\mathbf{Y}\mathbf{W}\mathbf{Y}^{\intercal}$, then a Slepian basis can be constructed from $\mathbf{D}$ by solving the Eigen problem $\mathbf{D}\mathbf{V}=\mathbf{V}\mathbf{\Lambda}$, where $\mathbf{V}$ is a matrix of Eigenvectors, each column of which forms a Slepian basis, and $\mathbf{\Lambda}$ is the diagonal matrix of Eigenvalues which quantify the concentration of each basis in the subregion $\mathcal{R}$. After sorting the Eigenvalues in descending order, the Shannon number $\left(P+1\right)^{2}A/4\pi$~\cite{wieczorek2007minimum} is then used to determine a cutoff rank based on the area $A$ of the subregion $\mathcal{R}$. 

\begin{figure}[t]
    \centering
    \subfloat[$\alpha=0$]{\includegraphics[width=0.32\linewidth]{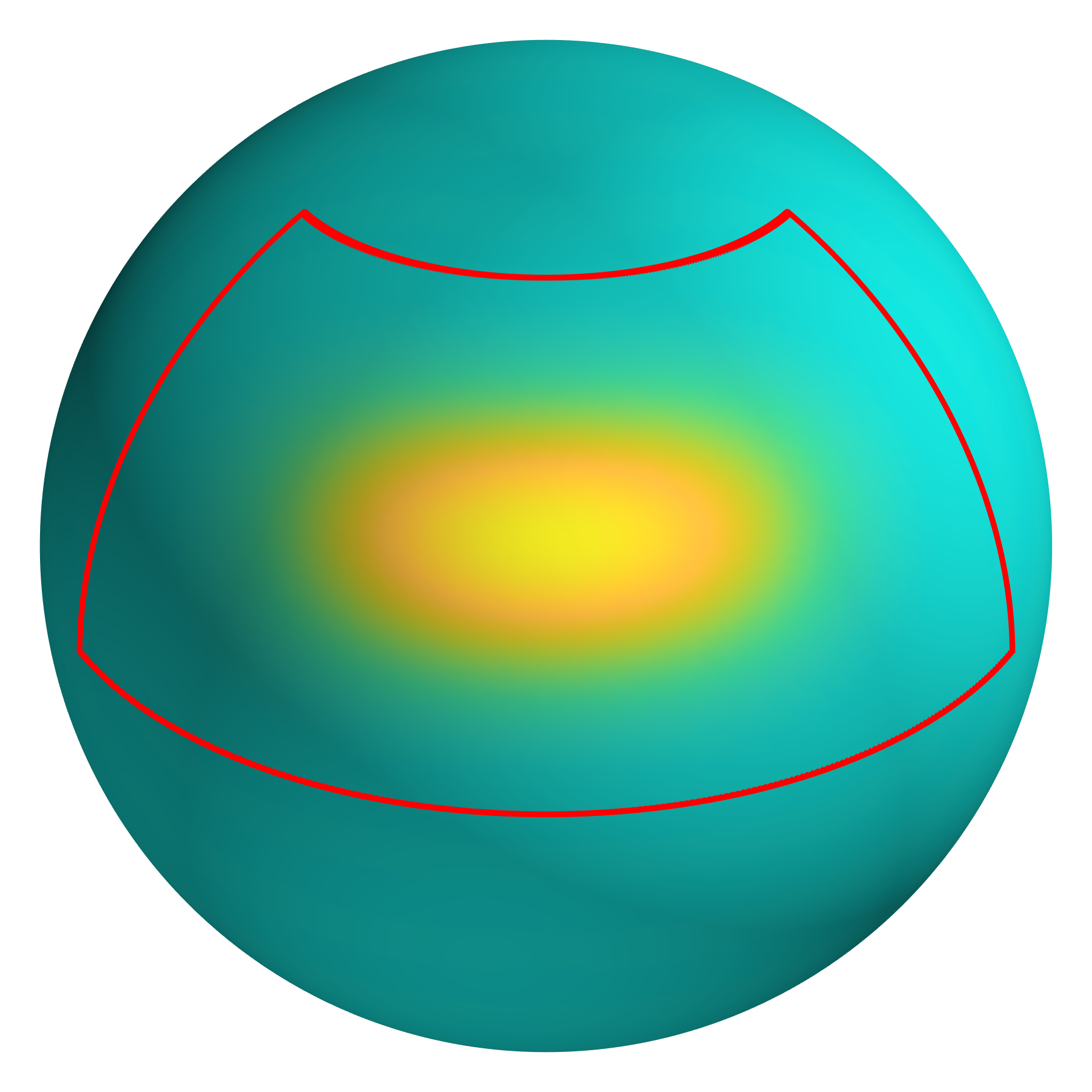}}
    \subfloat[$\alpha=6$]{\includegraphics[width=0.32\linewidth]{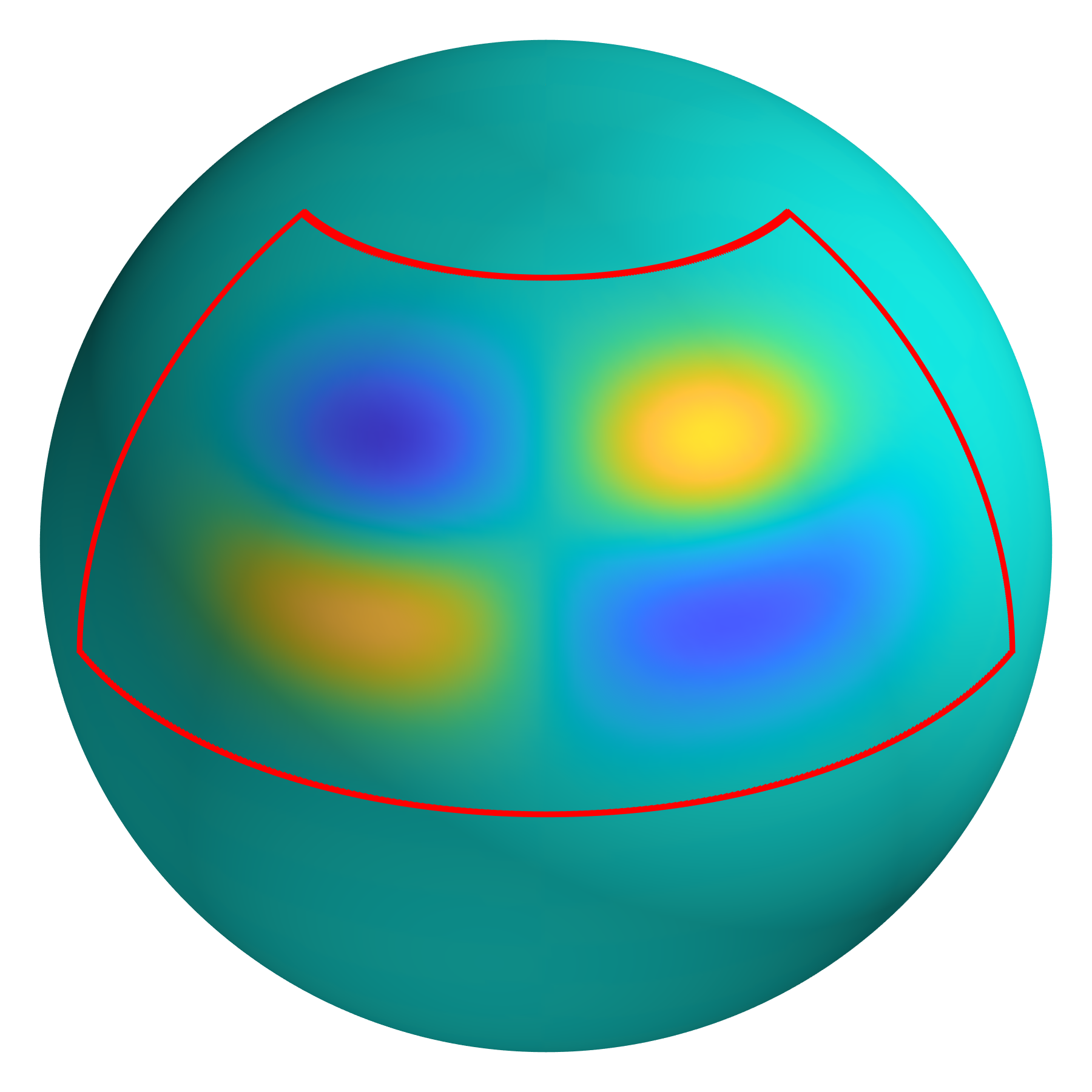}}
    \subfloat[$\alpha=10$]{\includegraphics[width=0.32\linewidth]{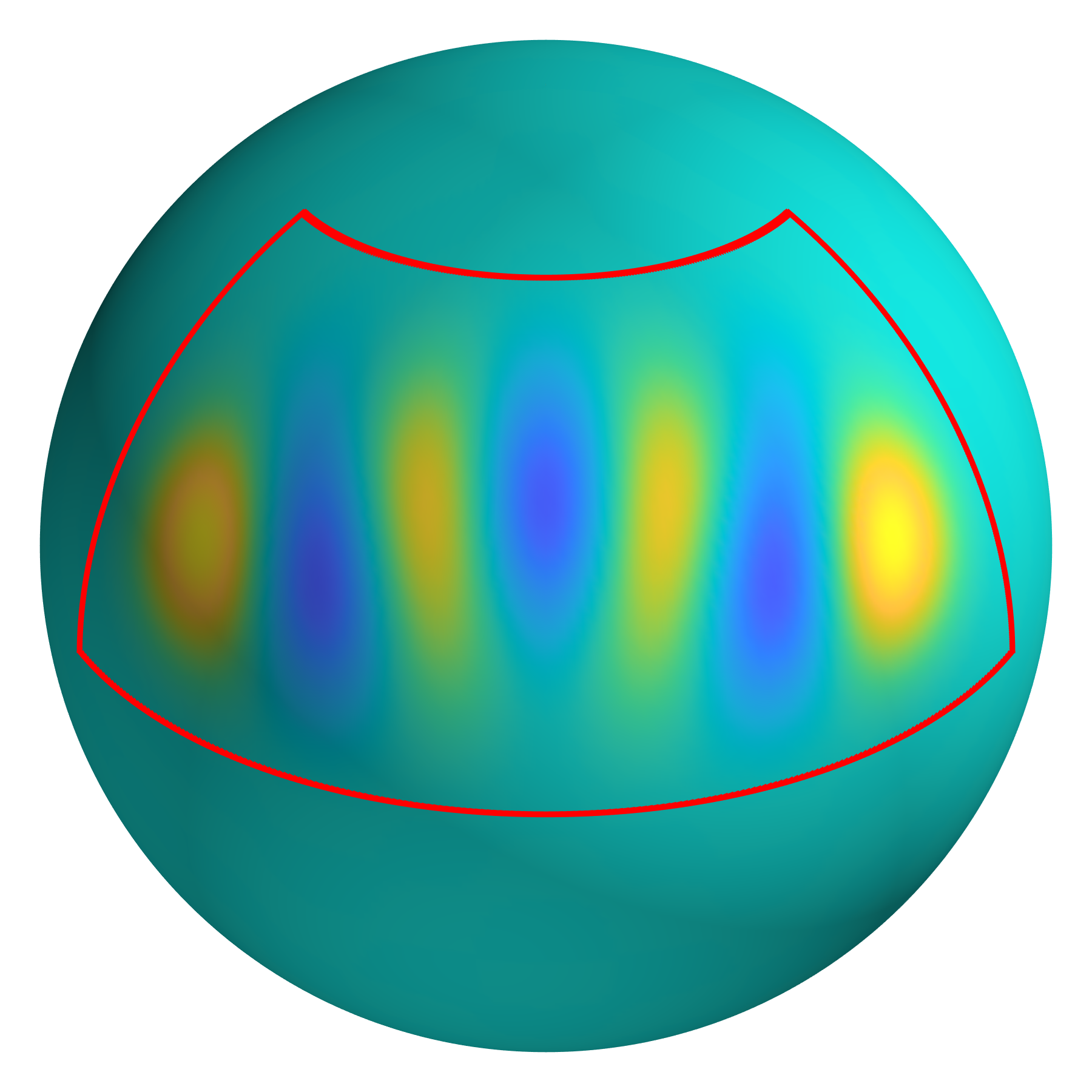}}
    \caption{Slepian spherical harmonic basis functions of varying rank $\alpha$, all well concentrated into a target subregion (red).}
    \label{fig:basis-functions}
\end{figure}

Figure~\ref{fig:basis-functions} shows three examples of tailored Slepian spherical harmonic basis functions, all are clearly concentrated in the red outlined target subregion. For the shown subregion, the maximum rank is $\alpha_{\text{max}} = 55$, making for 56 total basis functions. 

\subsection{Fitting Slepian basis to residual surface measurements}
Having defined the Slepian basis functions, we now need observations of the residual surfaces to estimate their weights. Given a scatterer of opportunity (or known target if available), in the field of view of the radar, there will be an associated peak in the radar data for each transmit-receive element pair. After processing, the unit-magnitude complex values at these peaks act as observations of the complex gains defined in~\eqref{eq:gains-only}. These observations can be converted to observations of the residual surfaces using the following normalization process. Consider the calibration process for one element and, for the sake of simplicity, we assume this is the 1st element in the transmit array. Thus, for the scatterer of opportunity, we have associated observations for transmit element 1 paired with each of the receive elements. Normalization is then performed to remove the influence of the receive elements by multiplying the observations with the complex conjugate of the observation corresponding to the reference transmit element at position 0 and that receive element. As a consequence, we can obtain $M_r$ measurements of $\Delta G_{t,1}$. 

If we now assume that in general $\ell$ is the element we are calibrating and $m$ is the index of the element in the array from which we are calculating~\eqref{eq:residual-vec} and $n$ is the observation index in the range $[0,N_s-1]$ for $N_s$ scatterers of opportunity/targets. Then the minimization problem for obtaining the weights of each Slepian basis is given by:
\begin{equation}\label{eq:min-prob}
   \min_{\left\{w_{a,\ell}\right\}_{a=0}^{\alpha_{\max}}} \sum_{m = 0}^{M_k-1} \sum_{n = 0}^{N_s-1} \left|\arg\!\left(\frac{g_{n,m,\ell}}{g_{n,m,0}}\right)-\sum_{a=0}^{\alpha_{max}}w_{a,\ell}\rho_{a,n}\right|^{2},
\end{equation}
where $\rho_{a,n}$ is the rank $a$ Slepian basis function evaluated in the direction of $n^\text{th}$ observation, $w_{a}$ is its corresponding weight and $g_{n,m,\ell}$ is the element of~\eqref{eq:gains-only} corresponding to elements $\ell$ and $m$ and with angles $\vartheta$ and $\varphi$ related to scatterer of opportunity $n$. We solve this minimization problem via least squares for each element. With this fitting method, all elements in the full MIMO array (inclusive of transmit and receive arrays) are calibrated.

\subsection{Comments}
Several assumptions are made in the derivation of our calibration method. The assumptions that the propagation channel influences the different MIMO waveforms equally, and that the arriving wavefront is planar are implicit in MIMO beamforming and are also assumed here. In addition, our calibration process assumes that the non-stationary channel is modeled well enough to accurately predict propagation paths. We anticipate that large HF antenna arrays will see the greatest benefit from our proposed calibration and the assumption of a well-modeled channel is common in calibration of OTHR and other HF radars~\cite{frazer2009mimo,chisham2021comparison,ponomarenko2015application}. Finally, further study of the influence of different noises on this calibration process is necessary. The noise present in a real channel may not be Gaussian, and due to the influence of the noise from different array elements in the calibration the process may benefit from a pre-whitening step when using real-world radar data.

\section{Analysis of Simulated Calibration Performance}
\label{sec:performance-analysis}
\begin{table*}[t]
    \centering
    \caption{Uncalibrated, $U$, and Calibrated, $C$, Beamformimg Performance}
    \label{tab:placeholder}
    \begin{tabular}{ccccccccccc}
    \toprule
        Observation SNR & Peak SNR Increase & Peak SNR Recovery & $\bar{\emptyset}_{U}$ dB & $\bar{\emptyset}_{C}$ dB & Uncal. $\Delta\theta$ & Cal $\Delta\theta$\\
    \midrule
        0   & 1.32 dB ($+35.6\%$) & $93.7\%$ & $-25$ & $-30.7$ & $2.7^{\circ}$ & $1.2^{\circ}$\\
        4   & 1.53 dB ($+42.4\%$) & $98.5\%$ & $-25$ & $-36.6$ & $2.7^{\circ}$ & $0.7^{\circ}$\\
        8   & 1.58 dB ($+44\%$) & $99.5\%$ & $-25$ & $-42.3$ & $2.7^{\circ}$ & $0.4^{\circ}$\\
        12  & 1.6 dB ($+44.4\%$) & $99.8\%$ & $-25$ & $-46$ & $2.7^{\circ}$ & $0.2^{\circ}$\\
    \bottomrule
    \end{tabular}
\end{table*}
\begin{figure}[t]
    \centering
    \includegraphics[width=0.99\linewidth]{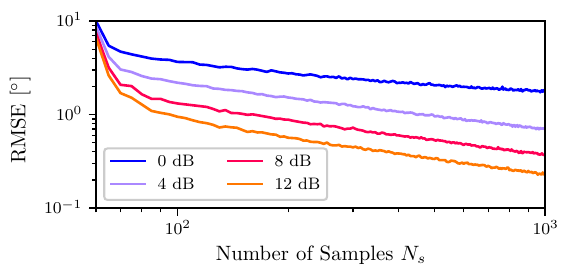}
    \caption{Relationship between number of samples and absolute difference between the estimated and ground truth Slepian weights.}
    \label{fig:snr-behaviour}
\end{figure}

To evaluate how well our method estimates the weights of the Slepian basis functions, we consider the problem of calibrating a MIMO array comprising 8 transmit and 8 receive elements. The residual surfaces for the elements are generated by randomly drawing the ground truth weights from a uniform distribution spanning $-0.05\leq w_{a,\ell}\leq0.05$ radians and the Slepian basis functions are concentrated in the region $-60^{\circ}\leq\vartheta\leq60^{\circ}$ and $5^{\circ}\leq\varphi\leq60^{\circ}$. Samples of these residual surfaces are distributed uniformly using Fibonacci nodes~\cite{hardin2016comparisonpopularpointconfigurations}. Noisy observations of~\eqref{eq:kronecker-beamformer} are simulated for each scatterer of opportunity, where complex-valued white Gaussian noise is added to produce SNR values of 0, 4, 8, and 12~dB. After normalization we solve the least squares problem and compare the obtained estimates of the weights with the ground truth. Figure~\ref{fig:snr-behaviour} shows the root mean square error (RMSE) between the ground truth and estimated weights, as a function of increasing number of scatterers $N_s$ and SNR. Each configuration of scatterer number $N_{s}$ and SNR was averaged over 50 runs, with each run using a new set of randomly generated Slepian weights. As expected, we observe that for the higher SNRs ($\geq4$ dB), increasing the number of scatterers improves the estimation accuracy of the weights, particularly for $N_{s}\gtrsim100$. The improvement in estimation accuracy for $0$ dB is less pronounced.
\begin{figure}[t]
    \centering
    \includegraphics[width=0.99\linewidth]{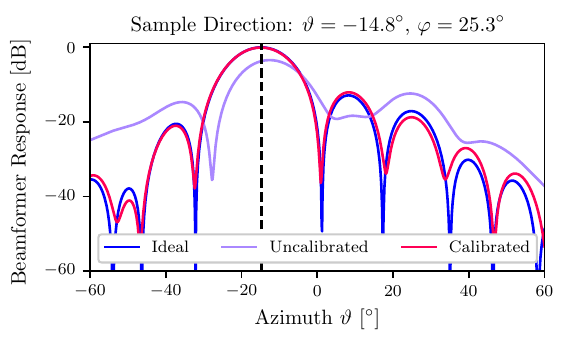}
    \caption{Beamformer output comparison for an ideal array (identical elements), a perturbed array, uncalibrated, and calibrated with a Slepian residual surface approximation computed from observations with an 8 dB SNR. Dashed black line indicates true azimuth.}
    \label{fig:beamformer-behaviour}
\end{figure}

Moving to beamforming performance, \autoref{fig:beamformer-behaviour} illustrates a slice in azimuth of the ideal, uncalibrated, and calibrated beam patterns of an 8 transmit and receive element MIMO array. The perturbations to the array were generated with the same uniform random weight method used for the Monte-Carlo simulation in \autoref{fig:snr-behaviour} with an SNR of 8~dB. The calibrated curve demonstrates the outcomes of calibration: the discrepancy between the direction of the realized main lobe and the ideal main lobe ($\Delta\theta$) is decreased; the peak of the main lobe is brought closer to the ideal level (Peak SNR Recovery); nulls are deepened ($\emptyset$ dB); and side-lobes are reduced. 

In \autoref{tab:placeholder}, the measures of beamformer quality described above are presented for scenarios with different observation SNRs. Results are averaged across 10 MIMO arrays with unique randomly generated Slepian weights, for 200 randomly sampled beamforming directions in each array's field of view. In general, the calibrated beamformer demonstrates increased output power (SNR Increase, \autoref{tab:placeholder}), bringing it within $>90\%$ of the ideal output power for the four scenarios in comparison to the uncalibrated beamformer (SNR Recovery, \autoref{tab:placeholder}). Similarly, nulls are deepened ($\emptyset_{U,C}$, \autoref{tab:placeholder}) and direction finding error improves for all calibrated beamformer output averages ($\Delta\theta$, \autoref{tab:placeholder}). The best performance occurs when the calibration is run with observations whose SNRs are $\geq$ 4 dB. For such cases, there is continued improvement in beamformer error, null depth, and SNR recovery. When calibration is completed with SNRs below this threshold, some improvement is still observed, but these are out of proportion with performance gains at higher observation SNRs. This disproportional behaviour is consistent with the trends observed in RMSE of Slepian coefficient estimation with observations $<4$ dB. Because observations for calibration may be taken from Doppler processed radar data, in addition to multiple measurements allowed by arrays on transmit/receive, these measurements enjoy the gains of traditional radar processing. Accordingly, the actual pulse-to-pulse SNR required for calibration is significantly lower than the SNRs in~\autoref{tab:placeholder}.


\section{Conclusion}
\label{sec:conclusion}
In this paper, we have demonstrated the calibration of MIMO arrays through the use of residual surfaces. These residual surfaces approximate the differences between a given element and a reference element in the array. Both transmit and receive calibrations are possible, with each array having its own reference element. We have shown that these residual surfaces can be generated from scatterers of opportunity using a least-squares fit. Because the observations for this fitting process come from peaks in each MIMO channel, multiple observations are available for each scatterer. Consequently, residual surfaces can be estimated to a high accuracy even when observations occur at lower ($\geq4$ dB) SNRs. We have also shown that the approximated residual surfaces can be used to correct for phase errors during the beamforming process. Applying these corrections during the beamforming process generates a beam pattern which is much more closely aligned with that of an ideal array, even when the observations used to approximate the residual surface had low SNRs.

\vfill\pagebreak
\bibliographystyle{IEEEbib}
\bibliography{ref}
\end{document}